\documentclass[referee]{aa} 
\usepackage{graphicx}
\usepackage{txfonts}
\usepackage{wasysym}
\begin{document}

   \title{Uncertainties in polarimetric 3D reconstructions \\
of coronal mass ejections}


\author{A. Bemporad\inst{1} \and P. Pagano\inst{2}}
\institute{{INAF-Osservatorio Astrofisico di Torino, via Osservatorio 20, 10025 Pino Torinese (TO), Italy.}
\and
{School of Mathematics and Statistics, University of St Andrews, North Haugh, St Andrews, Fife, Scotland KY16 9SS, UK.}
}

   \date{February 11, 2015}

 
  \abstract
   {}
   {The aim of this work is to quantify the uncertainties in the three-dimensional (3D) reconstruction of the location of coronal mass ejections (CMEs) obtained with the so-called polarization ratio technique. The method takes advantage of the different distributions along the line of sight of total ($tB$) and polarized ($pB$) brightnesses emitted by Thomson scattering to estimate the average location of the emitting plasma. This is particularly important to correctly identify of CME propagation angles and unprojected velocities, thus allowing better capabilities for space weather forecastings.}
   {To this end, we assumed two simple electron density distributions along the line of sight (a constant density and Gaussian density profiles) for a plasma blob and synthesized the expected $tB$ and $pB$ for different distances $z$ of the blob from the plane of the sky and different projected altitudes $\rho$. Reconstructed locations of the blob along the line of sight were thus compared with the real ones, allowing a precise determination of uncertainties in the method.}
   {Results show that, independently of the analytical density profile, when the blob is centered at a small distance from the plane of the sky (i.e. for limb CMEs) the distance from the plane of the sky starts to be significantly overestimated. Polarization ratio technique provides the line-of-sight position of the center of mass of what we call folded density distribution, given by reflecting and summing in front of the plane of the sky the fraction of density profile located behind that plane. On the other hand, when the blob is far from the plane of the sky, but with very small projected altitudes (i.e. for halo CMEs, $\rho<1.4$ R$_\odot$), the inferred distance from that plane is significantly underestimated. Better determination of the real blob position along the line of sight is given for intermediate locations, and in particular when the blob is centered at an angle of 20$^\circ$ from the plane of the sky.}
   {These result have important consequences not only for future 3D reconstruction of CMEs with polarization ratio technique, but also for the design of future coronagraphs aimed at providing a continuous monitoring of halo-CMEs for space weather prediction purposes.}


\titlerunning{Uncertainties in polarimetric 3D reconstructions of CMEs}
\authorrunning{Bemporad \& Pagano}

   \maketitle
%


\section{Introduction}

\begin{figure}[!thcb]
\begin{center}
\includegraphics[scale=0.5]{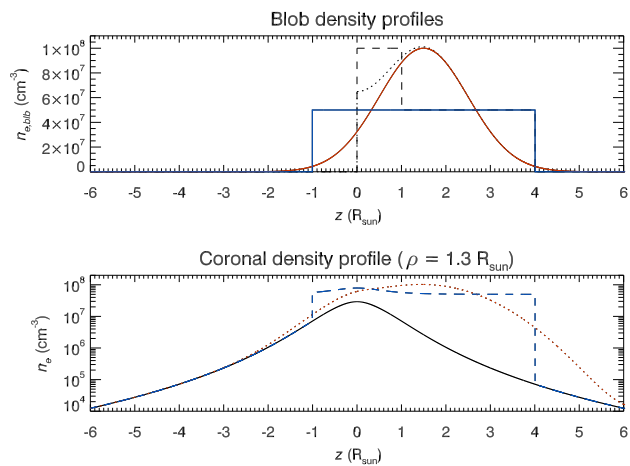}
\caption{
Top: blob density distribution $n_{e,blb}$ along the LOS coordinate $z$ for the case (A) of constant density (solid blue line) and the case (B) of gaussian density distribution (solid red line). For future reference this figure also shows the folded density distributions (see text) corresponding to the constant density blob (dashed line) and to the gaussian density blob (dotted line). Bottom: coronal density distribution $n_{e,cor}$ along the LOS (solid line) and total density distributions for the cases A (dashed blue line) and B (dotted red line).
}
\label{fig:fig00}
\end{center}
\end{figure}
Since the launch of twin STEREO spacecraft in 2006, the scientific community has devoted significant efforts to the development of data analysis techniques aimed at the three-dimensional (3D) reconstruction of plasma density distribution within coronal mass ejections (CMEs). Knowledge of the 3D CME structure is  crucial for many different reasons: first, it is the main ``driver behind the development of theoretical ideas'' \citep{Thernisien2011}, thus providing boundary conditions for CME models. Second, the determination of the 3D CME structure allows us to understand whether the event could interact with the Earth's magnetosphere; a better understanding of 3D CME evolution is needed in order to improve our capabilities of providing promp alerts for space weather forecastings. As a consequence, many different techniques for 3D reconstructions have been developed, such as triangulation via tie-pointing \citep{Inhester2006} or local correlation tracking \citep{Gissot2008}, forward modeling \citep{Thernisien2009}, inverse reconstruction \citep{Frazin2009}, constraints on the true CME mass \citep{Colaninno2009} and mask fitting \citep{Feng2012}. Various comparisons between these different methods when applied to the same event have also been performed by some authors for data provided by coronagraphs \citep[see e.g.][]{Mierla2010,Feng2013} and heliospheric imagers \citep[e.g.][]{Mishra2014}. In general it was found that the CME propagation direction can be determined with all these methods within an uncertainty of $\sim 10^\circ$, but derived 3D spatial extensions of CMEs are not fully consistent with each other; moreover, reconstructions using three-view observations are more precise than those made with only two views and different methods have advantages or disadvantages depending on the angular distance between STEREO and other spacecraft. In addition, CME arrival times predicted at 1 AU with different methods still have a large uncertainty (around 10--30 hours).

All the methods mentioned above deal with the analysis of images acquired at the same time by multiple spacecraft. Nevertheless, before the launch of STEREO and the availability of this kind of multiple-view-point observations, a very promising method for the study of 3D distribution of CME plasma with single-view-point images was published by \citet{Moran2004}. This method is mainly based on the dependency of Thomson scattering on the scattering angle, hence on the location $z$ of the electrons along the line of sight (LOS), with the property that the polarized ($pB$) and unpolarized ($uB$) brightnesses have a slightly different dependence on this angle. This allows the determination of the average plasma location along the LOS from the $pB/uB$ ratio observed in single view-point images, with a well known $\pm z$ ambiguity due to the symmetry of Thomson scattering about the plane of the sky (POS; $z = 0$). The $pB/uB$ ratio can be computed pixel by pixel in the 2D coronagraphic image, thus providing a 3D cloud of points, each point representing the location along $z$ of CME plasma with some kind of unknown LOS averaging. This technique, usually referred to as the polarization-ratio technique, was also validated in the STEREO era \citep{Mierla2010,Moran2010} with comparisons between 3D reconstructions obtained with polarization measurements and other reconstruction methods. Very recently a classification of possible ambiguities arising from polarimetric reconstructions of CMEs depending on the location of CME structures with respect to the POS was given by \citet{Dai2014}.

The present work aims at providing a key with which to correctly interpret results and to estimate quantitatively the uncertainties in polarization ratio technique. This will be done by assuming two simple density distributions of a plasma blob along the LOS, superposed over a typical coronal density distribution, by synthesizing the corresponding $pB$ and $uB$ emissions, and by comparing the LOS blob location derived with polarization ratio with its real location. Results from this analysis will be very important in particular for the future METIS coronagraph \citep{Antonucci2012,Fineschi2013} that will provide polarized white light images from the unique vantage point offered by the Solar Orbiter spacecraft.
\begin{figure*}[!thcb]
\begin{center}
\includegraphics[scale=1.0]{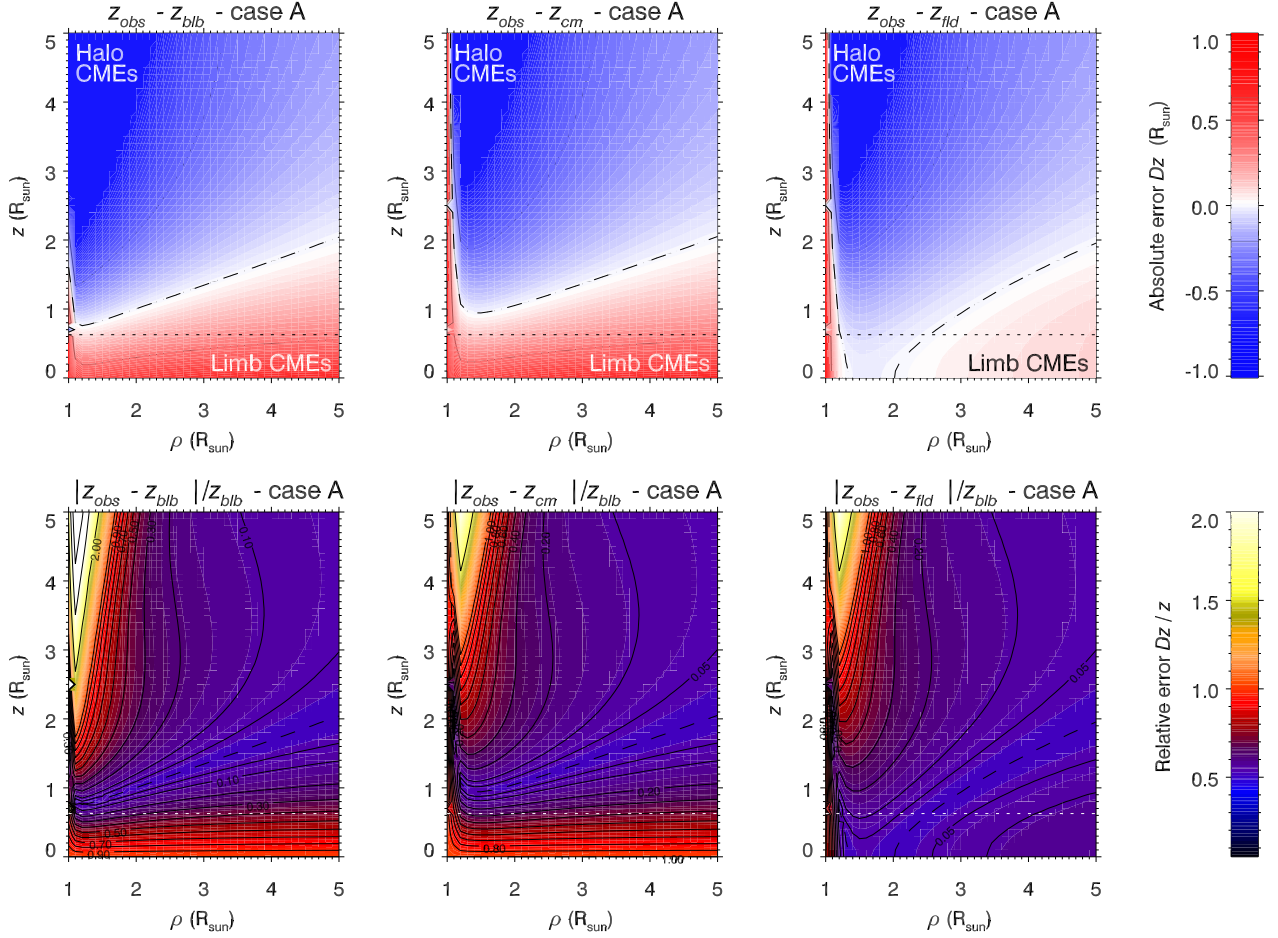}
\caption{
Top row: errors in the determination of a CME position along the line of sight as a function of the projected position on the plane of the sky ($\rho$, $x-$axis) and the distance from that plane ($z$, $y-$axis) under the hypothesis of a constant density distribution across the CME (case A). The plots show regions where the distance is under (over) -estimated in blue (red) with respect to the real location of the CME (left), the center of mass (middle), and the folded density center of mass (right, see text). Bottom row: same as in the top row showing the relative errors. In all the plots the dashed black line shows the points where the location of the blob is determined without errors, while horizontal white dotted line marks the region where the blob distance from the plane of the sky starts to be smaller than $1/4$ of the half width of the blob ($z_{blb}<\sigma_{blb}\sqrt{2\pi}/4$).
}
\label{fig:fig01}
\end{center}
\end{figure*}

\section{Uncertainties in single blob reconstruction}
\label{blob}

The polarization ratio technique has been extensively described by previous authors and very nicely reviewed by \citet{Dai2014}, who also gave again the explicit equations for the total ($tB$) and polarized ($pB$) brightnesses. Here we simply note that, given these equations, the ratio between $pB$ and $tB$ emitted by a single scattering electron along the LOS is given by
\begin{equation}
\frac{pB_z}{tB_z} = \frac{\left[ (1-u)A+uB  \right] (1-z^2/r^2)}{2\left[ (1-u)C+uD \right]-\left[(1-u)A+uB\right] (1-z^2/r^2)},
\end{equation}
where $u$ is the limb darkening coefficient in the visible wavelength of interest; $A, B, C,$ and $D$ are all geometrical functions depending only on the solid angle $\Omega$ subtended by the solar disk at the scattering electron location $z$ along the LOS; and $r$ is the heliocentric distance of this point \citep[see][for the explicit expression of these geometrical functions]{Altschuler1972}. The LOS coordinate is $z = 0$ on the POS and the above ratio depends on $z^2$, hence two points located symmetrically with respect to the POS at $\pm z$ correspond to the same value of this ratio. We note that this ratio is independent of the value of the local electron density $n_e$, but the observed ratio will be computed between the total (i.e. integrated along the whole LOS) $pB$ and $tB$ quantities, which are both dependent on the (in principle unknown) LOS density distribution $n_e(z)$. In the integration along the LOS the same $n_e(z)$ distribution is weighted with different geometrical functions $A, B, C, D$ to give the observed $pB$ and $tB$.
Hence, if there is a spatially limited region where the density is locally much larger along the LOS (as happens during CMEs), the contribution to the total ratio is dominated by the emission from this region.
Thus, it is possible to derive an average $\langle z^2 \rangle_{n_e}$ which is a good approximation for the real location of the emitting region along the LOS.

\begin{figure*}[!htcb]
\begin{center}
\includegraphics[scale=1.0]{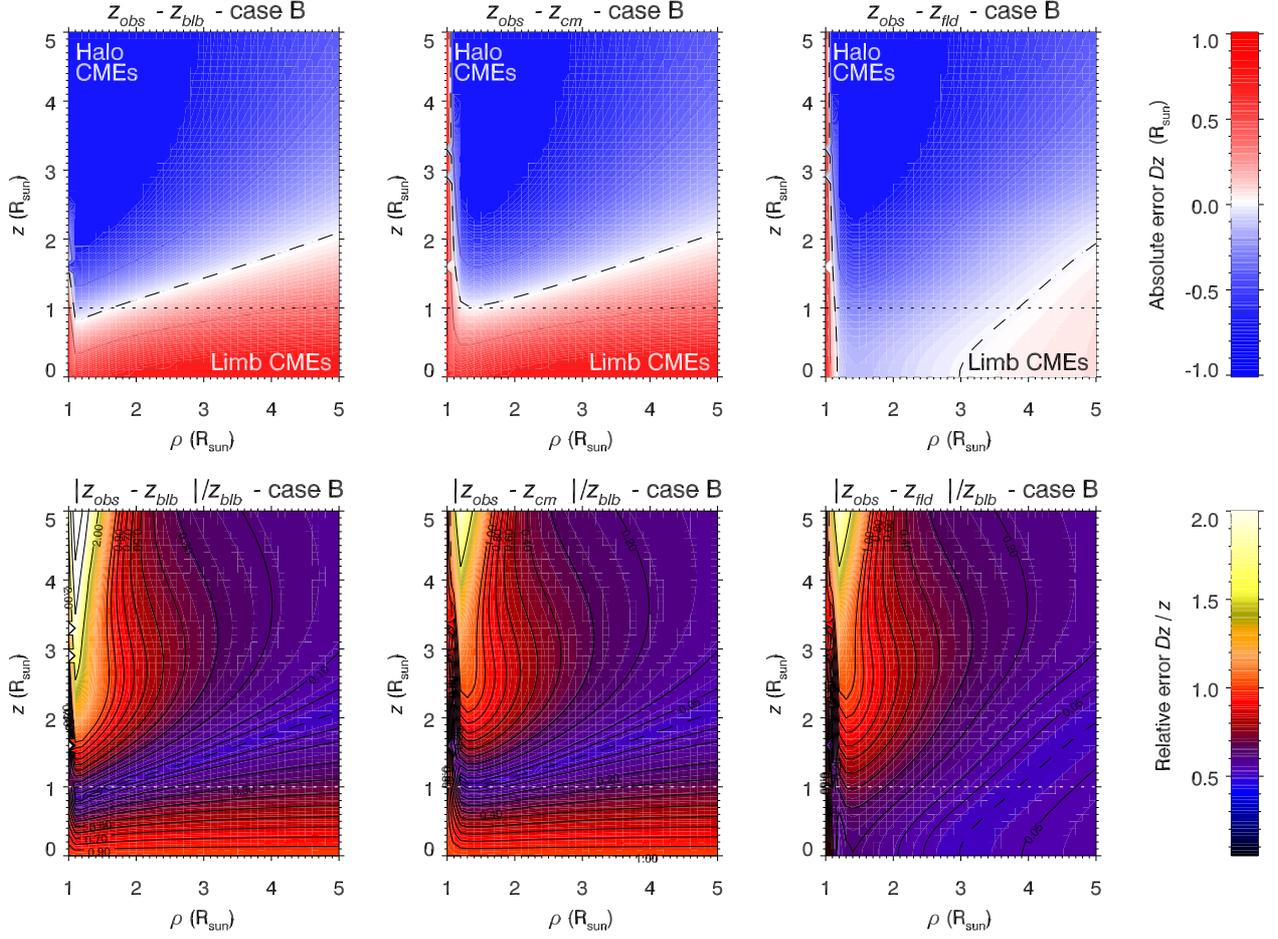}
\caption{
Top row: errors in the determination of a CME position along the line of sight as a function of the projected position on the plane of the sky ($\rho$, $x-$axis) and the distance from that plane ($z$, $y-$axis) under the hypothesis of a Gaussian density distribution across the CME (case B). The plots show regions where the distance is under (over) -estimated in blue (red) with respect to the real location of the CME (left), the center of mass (middle), and the folded density center of mass (right, see text). Bottom row: same as in the top row showing the relative errors: dashed line show the ideal location of the CME where the error is zero. In all the plots the dashed black line shows the points where the location of the blob is determined without errors, while horizontal dotted line marks the region where the blob distance from the plane of the sky starts to be smaller than the $1/e$ half width of the blob ($z_{blb}<\sigma_{blb}$).
}
\label{fig:fig02}
\end{center}
\end{figure*}

\subsection{Description of the technique}

The aim of the present work is to characterize the correctness of this approximation and the uncertainties related with the polarization ratio technique. We will focus here on simple analytical density profiles along the LOS, while more realistic density distribution obtained with a magneto-hydrodynamic (MHD) simulation will be analyzed in the near future with a second work. To this end, we investigate the reconstruction of the real location along the LOS of known spatially limited plasma density distributions, representing an erupting feature (like a CME). In particular, we consider along the LOS coordinate $z$ a single unidimensional plasma blob with two possible distributions of the electrons density: 1) a constant density $n_{e,blb} = n_{e,blb0}$ within a fixed $z$ interval (case A), and 2) a Gaussian density distribution $n_{e,blb} = n_{e0} \exp{\left[-(z-z_{blb})^2/(2\sigma_{blb}^2)\right]}$. Both distributions (shown in the top panel of Figure \ref{fig:fig00}) are centered at the distance $z_{blb}$ from the POS, and have LOS extensions defined by the $1/e-$half width $\sigma_{blb}$ and density peak $n_{e0}$ for the Gaussian density distribution, and half width $\sigma_{blb}\sqrt{2\pi}$ and density $n_{e0}/2$ for the constant density distribution. We selected two different density distributions inside the blob to demonstrate that this choice does not significantly affect the results presented here. We also note that a Gaussian density distribution is at the base of the well-known graduated cylindrical shell (GCS) model proposed by \citet[][]{Thernisien2009} and employed by many authors to perform 3D forward modeling of CME observed by STEREO and SOHO coronagraphs. In particular, the CGS model assumes the flux rope to have the shape of a tubular hollow croissant, with electron density placed on the shell by using a Gaussian-like distribution function \citep[see Eq. 3 in][]{Thernisien2009}. Hence, a single LOS passing across the shell of the GCS model meets two times a Gaussian-like distribution function similar to the blob simulated here, the first time at the edge of the shell located towards the observer, and the second one at the edge of the shell located behind.

The considered LOS is placed at the projected distance $\rho$ from the center of the Sun. We also assume from the literature an electron density distribution $n_{e,cor}$ \citep[Eq. 3 in][]{Gibson1999} for the surrounding corona aligned along the LOS (shown in the bottom panel of Figure \ref{fig:fig00}, solid black line) and verified that the selection of this curve does not affect the results presented here. The density at each point $z$ along the LOS is simply given by $n_e(z)=n_{e,cor}(z) + n_{e,blb}(z)$ (bottom panel of Figure \ref{fig:fig00}). It should be noted that whereas $n_{e,cor}(z)$ always peaks at the POS, the peak of the $n_{e,blb}(z)$ is obviously located at position $z_{blb}$ and the resulting total distribution $n_e(z)$ has a profile in general not symmetric around any value of $z$. We then synthesize the total ($tB$) and polarized ($pB$) brightnesses in white light obtained by varying the projected distance $\rho$ of the LOS from 1 R$_\odot$ to 5 R$_\odot$ and the LOS location $z_{blb}$ of the blob center from 0 R$_\odot$ to 5 R$_\odot$, both with steps of 0.1 R$_\odot$. The ratio between synthesized $pB_{obs}$ and $tB_{obs}$ integrated along the LOS was then computed as
\begin{equation}
\frac{pB_{obs}}{tB_{obs}}=\frac{\int_{LOS} pB_z [n_{e,cor}(z) + n_{e,blb}(z)]\, dz}{\int_{LOS} tB_z [n_{e,cor}(z) + n_{e,blb}(z)]\, dz}.
\end{equation}
This ratio was then compared with the theoretical ratio (Eq. 1), and the observed blob distance $z_{obs}$ from the POS was determined as the distance where the observed and theoretical ratios are the same \citep[see][for more details]{Moran2004}. This computation was performed for each LOS $\rho$ and each location of the blob $z_{blb}$ and the resulting $z_{obs}$ values were compared with the assumed ones. 

\begin{figure*}[!thcb]
\begin{center}
\includegraphics[scale=1.0]{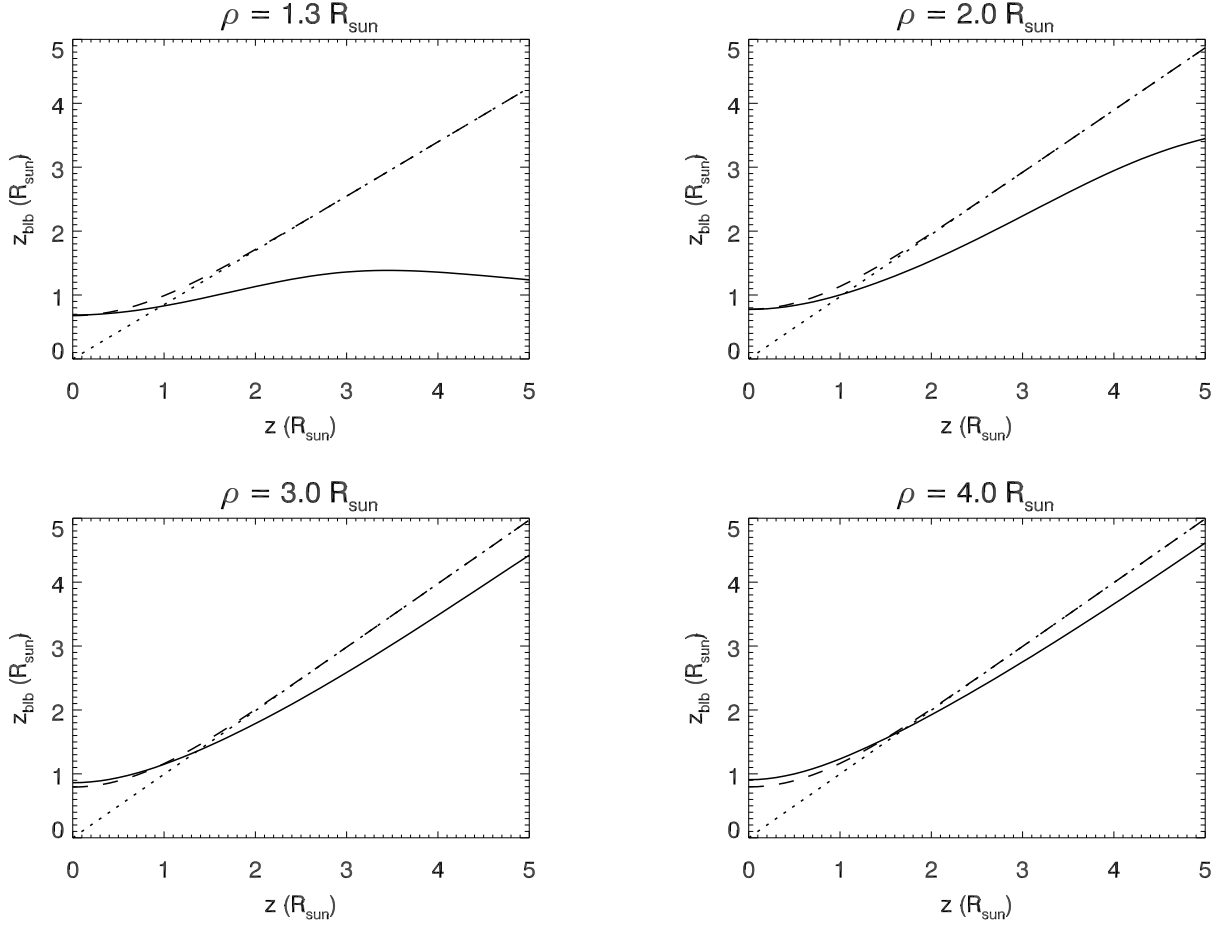}
\caption{
Comparison at four different LOS distances $\rho$ between the real location $z_{blb}$ of the blob (for case B) along the LOS (dotted line), the location $z_{obs}$ inferred from polarization ratio technique (solid line), and the location $z_{fld}$ of the center of mass for the folded density distribution case (dashed line).
}
\label{fig:fig03}
\end{center}
\end{figure*}

\subsection{Results: limb CMEs}

The resulting distributions of the quantities $z_{obs}-z_{blb}$ (top panels) and $|z_{obs}-z_{blb}|/z_{blb}$ (bottom panels) are shown in Figures \ref{fig:fig01} and \ref{fig:fig02} for the case A and case B blob, respectively. In each panel these quantities are shown for different LOS projected distances ($\rho$, $x-$axis) and different locations ($z$, $y-$axis) of the unidimensional blob along the LOS, as we obtained by assuming a $1/e-$half width $\sigma_{blb}=1$ R$_\odot$ with density peak $n_{e0}=10^8$ cm$^{-3}$ \citep[equal to the coronal density at the heliocentric distance of 1.17 R$_\odot$ with the][profile]{Gibson1999} for the Gaussian density blob (case B), and half width $\sigma_{blb}\sqrt{2\pi} \simeq 2.5$ R$_\odot$ with density $n_{e0}/2 = 5 \times 10^7$ cm$^{-3}$ for the constant density blob (case A). First of all, results in both Figures \ref{fig:fig01} and \ref{fig:fig02} show that when the blob is very close to the POS (i.e. for $z_{blb} < \sigma_{blb}\sqrt{2\pi}/4 \simeq 0.63$ R$_\odot$ and  $z_{blb} < \sigma_{blb} = 1$ R$_\odot$ for case A and case B, respectively) the blob distance from that plane is significantly overestimated: this is the case of limb CMEs. As we verified by increasing the LOS extension of the blob up to $\sigma_{blb} = 2$ R$_\odot$, the above results remain basically unchanged. Moreover, comparison between panels in Figures \ref{fig:fig01} and \ref{fig:fig02} also shows that the above results are not significantly affected by the LOS density distribution inside the blob.

A second interesting result is that the blob positions derived from the polarization ratio technique corresponds basically to the location of the plasma center of mass along the considered LOS. This is shown in the middle panels of Figures \ref{fig:fig01} and \ref{fig:fig02}, where we plot the distribution of the quantities $z_{obs}-z_{cm}$ (top) and $|z_{obs}-z_{cm}|/z_{blb}$ (bottom). A comparison between the middle and left panels of both figures shows that no significant difference exists: this means that (at least for the LOS density distributions assumed in the present work) polarization ratio technique provides a quite good approximation of the location along the LOS of the center of mass of total (i.e. coronal plus CME) density distribution (provided that the blob is propagating not too close to the POS and not too close to the solar limb). The reason for this is simply that the position of the center of the blob is almost coincident with the position of the center of mass of total density distribution along the LOS.

\begin{figure*}[!thcb]
\begin{center}
\includegraphics[scale=1.0]{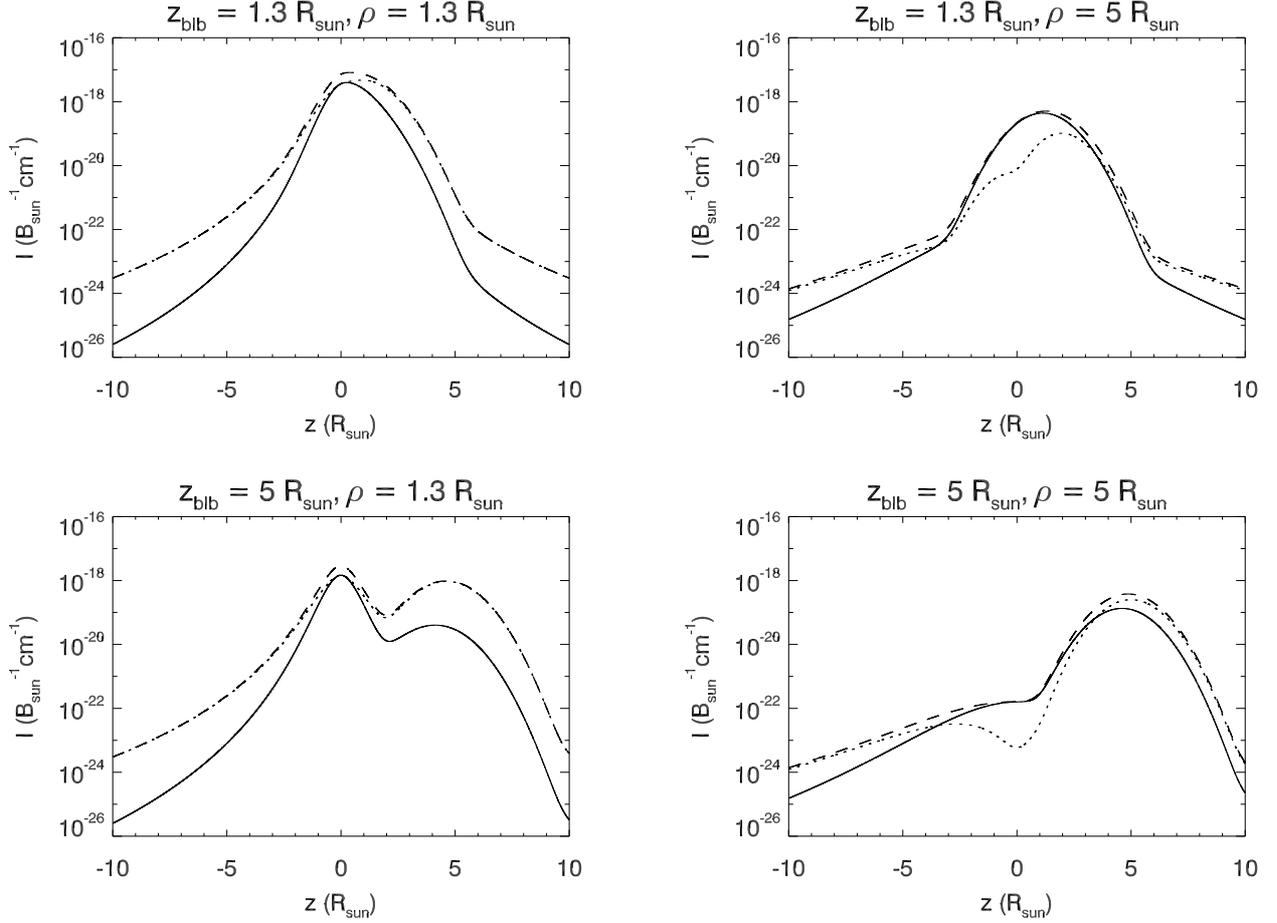}
\caption{
Distributions along the LOS of the polarized $pB$ (solid line), unpolarized $uB = tB-pB$ (dotted line) and total $tB$ (dashed line) visible light brightnesses for two different lines of sight ($\rho = 1.3$ R$_\odot$, left column, and $\rho = 5$ R$_\odot$, right column) and two different distances of the blob from the POS ($z_{blb} = 1.3$ R$_\odot$, top row, and $z_{blb} = 5$ R$_\odot$, bottom row).
}
\label{fig:fig04}
\end{center}
\end{figure*}
The large disagreement in the 3D location of the blob when it is very close to the POS occurs because the polarization ratio technique provides the position along the LOS with a $\pm z$ ambiguity, due to the simmetry of Thomson scattering about $90^\circ$: two electrons placed at $z$ and $-z$ cannot be distinguished with the polarimetric technique. Hence, as soon as a significant part of the blob starts to be located behind the POS, the technique gives a $z$ coordinate which is a mixture between the plasma located in front of and behind that plane. In order to quantitatively show this point we changed the blob density distribution by cutting the density located behind the POS ($z<0$) and by summing it back (reflected about $z=0$) over the part of the blob located in front of that plane ($z>0$): for clear reasons we call this ``folded'' blob density distribution (see Figure \ref{fig:fig00}). For each location of the blob in the $(\rho,z)$ plane we thus determined the new position of the center of mass $z_{fld}$ of the folded density distribution and compared it with the blob location as inferred the from polarization ratio technique, for both cases A and B. By plotting the quantities $z_{obs}-z_{fld}$ (top right panels in Figure \ref{fig:fig01} and Figure \ref{fig:fig02}) and $|z_{obs}-z_{fld}|/z_{blb}$ (bottom right panels in Figure \ref{fig:fig01} and Figure \ref{fig:fig02}) it is clear that the error in the location of the blob close to the POS is removed. All these results hold for both cases A and B; in order to simplify the discussion, in what follows we simply refer to results obtained for the Gaussian density distribution inside the blob (case B).

The above result means that the polarization ratio technique, when applied to limb CMEs, provides in first approximation the location of the center of mass of the coronal plus CME folded density distribution; this clearly gives an overestimate of the real CM distance from the POS (\ref{fig:fig02}). In order to better show this point, Figure \ref{fig:fig03} shows a comparison between the real ($z_{blb}$, dotted line) and the derived ($z_{obs}$, solid line) blob positions (case B), together with the location of the center of mass computed with the folded density distribution (dashed line). The panels in Figure \ref{fig:fig03} show that when $\rho \apprge 2$ R$_\odot$ then $z_{obs} \approx z_{fld}$. Small differences between $z_{obs}$ and $z_{fld}$ are present everywhere: closer to the POS ($z<1$ R$_{\odot}$) is $z_{obs} > z_{fld}$, while farther from that plane we find that systematically $z_{obs} < z_{fld}$ for any projected distance $\rho$ of the LOS.

\subsection{Results: halo CMEs}

The top panels of Figure \ref{fig:fig01} (case A) and Figure \ref{fig:fig02} (case B) also show that when the blob is expanding much farther from the POS (i.e. in a narrow cone pointing toward the observer), the measured distance is significantly understimated: this is the case of the halo CMEs. Very interestingly, when the blob is expanding in the intermediate regions, hence not too close to POS and not too close to the LOS, the uncertainty in the estimated position of the blob is in general quite small (not larger than $\sim 20$\%, bottom left panel of Figures \ref{fig:fig01} and \ref{fig:fig02}). Hence, the polarization ratio technique can provide good results even for halo CMEs expanding towards the observer along lines of sight not closer than $\rho \sim 1.3$ R$_\odot$ to the sun's center, so for the case of partial-halo CMEs. This makes information derived with the polarization ratio technique - for instance - with data acquired by the SOHO/LASCO-C2 coronagraph quite accurate even for partial halo CMEs, when a single CME dense feature is aligned along the LOS.

The explanation why the distance from the POS is significantly underestimated when the blob expands towards the observer with very small $\rho$ projected distances (blue regions in top panels of Figure \ref{fig:fig01} and Figure \ref{fig:fig02}) is completely different and mainly depends on Thomson scattering geometry: the same electron gives very different relative contributions to the polarized $pB$ and unpolarized $uB = tB - pB$ brightnesses just depending on its position along the LOS, hence depending on the value of the scattering angle $\chi$ between the radial pointing from the Sun to the electron position and the line connecting this point with the observer. In particular, the polarized brightness $pB$ emitted by the plasma volume in the scattering region at the location $z$ along the LOS is proportional to $\sin^2 \chi = (\rho/r)^2=\rho^2/(z^2+\rho^2)$, a quantity that maximizes on the POS where $z=0$. On the other hand, the total brightness $tB$ is proportional to the quantity $(1+\cos^2\chi)$, hence the degree of polarization $pB/tB$ changes along the LOS as $\sin^2\chi/(1+\cos^2\chi) = \rho^2/(\rho^2+2z^2)$, with the maximum $pB/tB$ on the POS. This also corresponds to the $pB$ distribution along the LOS being in general much more confined to the region closer to the plane of sky, while the $tB$ distribution is much broader. This results in a ratio $pB/uB$ emitted per electron which is significantly peaked on that plane, independently of the electron density distribution. This is in fact what allows the determination of the blob location along the LOS with the polarization ratio technique.

\begin{figure*}[!thcb]
\begin{center}
\includegraphics[scale=1.0]{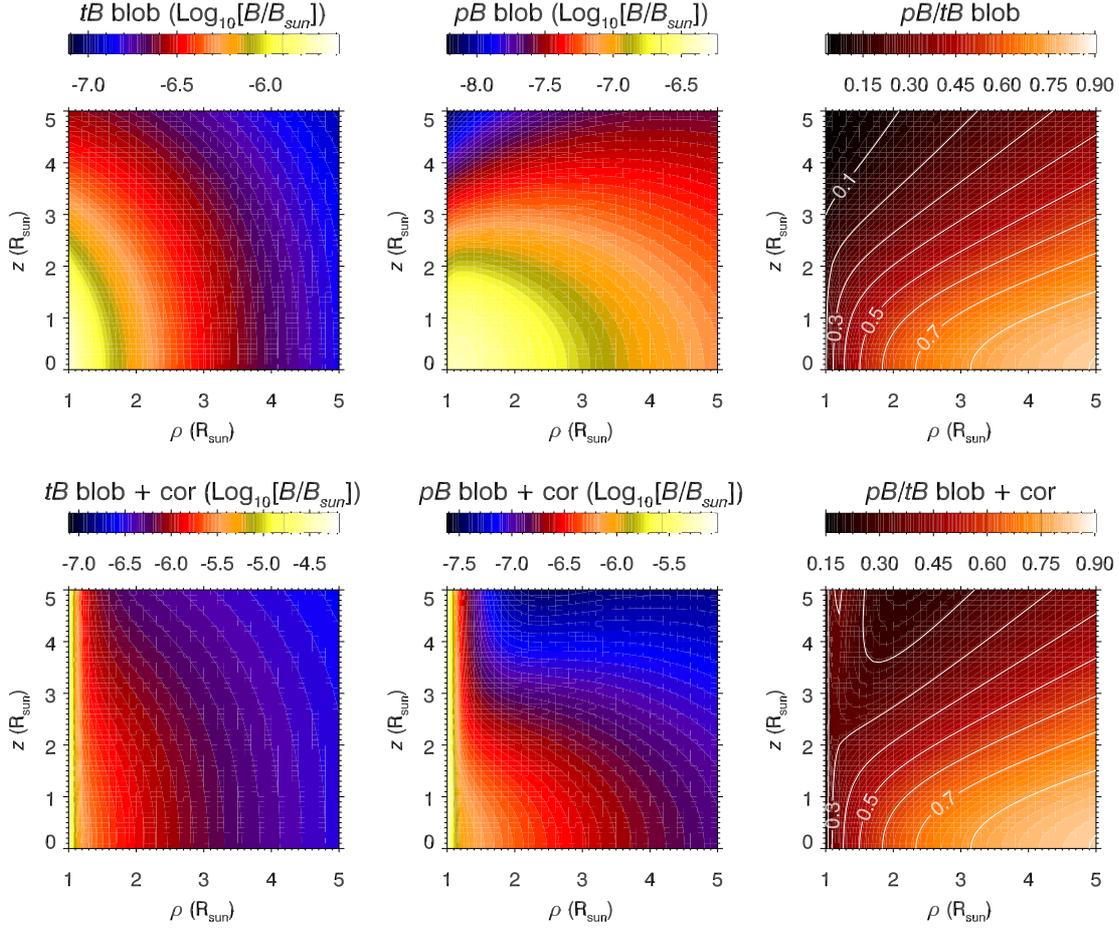}
\caption{
Top: distribution of the total ($tB$, left) and polarized ($pB$, middle) brightnesses emitted by the blob integrated along the LOS by assuming different positions for the blob on the $(\rho,z)$ plane, together with the relative $pB/tB$ intensity ratio (right). Bottom: same as in the top row for the total (i.e. corona plus blob) emissions.
}
\label{fig:fig05}
\end{center}
\end{figure*}
Nevertheless, depending on the considered $\rho$ value, different scattering angles are spanned within the same LOS extension ($z = \pm 10$ R$_\odot$ in our computation), leading to different values of the $pB/uB$ emitted per electron. For instance, along the LOS at $\rho = 5$ R$_\odot$ the scattering angle goes from $\chi=90^\circ$ to a minimum of $\arctan(\rho/z) \simeq 26.6^\circ$, while along the LOS at $\rho = 1.3$ the scattering angle goes down to $\arctan(\rho/z)\simeq 7.4^\circ$. As a consequence, along the LOS at $\rho = 5$ R$_\odot$ the $pB/uB$ per electron is $\sim 15-25$ times larger than the same ratio along the LOS at $\rho = 1.3$ R$_\odot$. For this reason, when the projected blob distance $\rho$ is large enough (right column in Figure \ref{fig:fig04}), the emitted $pB$ (solid line) is a significant fraction of $tB$ (dashed line) both when the blob is close (top right) or far (bottom right) from the POS. On the contrary, when the projected blob distance is small (left column in Figure \ref{fig:fig04}), the ratio $pB/uB$ is much smaller almost everywhere along the LOS and a significant $pB$ emission with respect to $tB$ is recovered only when the blob is closer to the POS (top left), while $pB \ll tB$ across the blob in the other case (bottom left). This will be pointed out again in the Conclusions and further discussed with Figure \ref{fig:fig05} (see below).  

Moreover, the region in the $(\rho,z)$ where the blob distance from the POS is understimated (blue regions in top panels of Figure \ref{fig:fig01} and Figure \ref{fig:fig02}) is very extended for another reason. The blob (as every CME) has a significant extension along the LOS coordinate $z$ ($\sigma_{blb}=1$ R$_\odot$ and $\sigma_{blb}\sqrt{2\pi}=2.5$ R$_\odot$, respectively for case B and A) and this creates an imbalance in the $pB$ emission between the half blob located closer to and the one farther from the POS. The emission coming from the latter part is less polarized than the emission coming from the former part, leading to an imbalance in the $pB/uB$ ratio along the LOS which (once the integration along the LOS is performed) turns out to be higher than the value of the theoretical ratio corresponding to the real position of the blob center, hence leading to an underestimate of the blob distance from the POS.

\section{Discussion and conclusions}
\label{conclusion}

The present work aims to explain in detail how the results of polarimetric imaging have to be interpreted in order to properly reconstruct the 3D structure of CMEs. To this end, polarized $pB$ and total $tB$ white light brightnesses of a plasma blob have been synthesized by assuming two simple blob density distributions (a constant and a Gaussian distribution) superposed onto an external coronal density distribution taken from the literature \citep{Gibson1999}. In simple terms we found that for both line of sight (LOS) blob density distributions the polarization ratio technique will overestimate the real distance from the plane of the sky (POS) when the observed structure propagates close to that plane (red regions in top panels of Figure \ref{fig:fig01}), hence for limb CMEs. This result can be understood by defining a folded density distribution, where for the fraction of the blob located behind the POS ($z<0$) the density distribution is reflected about $z=0$ and summed over the density profile in front of that plane ($z>0$, see Fig. \ref{fig:fig02}). It turns out that the 3D cloud of points resulting from the polarization ratio technique applied pixel by pixel to 2D real images of CMEs represents the location of the center of mass of the folded density distribution along the LOS. The CME fraction partially located behind the POS contributes to the center of mass distribution of the fraction located in front of that plane. Hence, care should be taken for limb events, and analysis of data acquired by more spacecraft than one is likely required, as recently pointed out by \citet{Dai2014}.

On the other hand, for CMEs originating near the center of the disk (like halo CMEs) we expect very little of the ejected plasma to cross the POS, and the polarimetric imaging technique is likely to properly describe the CME geometry when one wants to estimate the position along the LOS of a single structure (e.g. the ejected flux rope, the tip of the front), unless several structures are aligned along the same LOS. The latter case presents a limit in particular when one wants to visualize the 3D structure of the shock and the front, which are expected to have a significant extension along the LOS, while determination of bright and compact features (like CME cores) should be more precise. Nevertheless, as pointed out by \citet{Mierla2011}, this is true as long as the H$\alpha$ emission is not included in the band-pass of the coronagraphs. When this emission is included (as for STEREO coronagraphs) H$\alpha$ produces a strong spurious emission which, because it is not due to Thomson scattering by free electrons, provides unreliable results when the polarization ratio technique is applied to the prominence plasma often embedded in CME cores. At the same time, when the 3D position of the emitting prominence plasma can be determined via triangulation, a combination with the polarization ratio technique allows one to distinguish between the polarization due to Thomson scattering and H$\alpha$ polarized emission, as recently shown by \citet{dolei2014}.

Observations of structures far from the POS are also affected by another uncertainty due to the different brightness in polarized light of the plasma located along the LOS farther from that plane. Unlike previously, this effect leads to an underestimate of the distance from that plane of the density structure. The effect of distance overestimate close to the POS and underestimate far from it balance out at about $20^{\circ}$ from that plane, and above a projected LOS distance $\rho \apprge 1.3$ R$_{\odot}$, where the errors are of a few tenths of solar radii. Hence, in some conditions the polarization ratio technique is extremely accurate and reliable: the conditions are that only one structure should present along the LOS and that it should be far enough from the POS and from the projected location of solar limb. The method gives its best performance when the angle between the observed structure and the POS is about $20^{\circ}$, while far from these conditions we should estimate an error bar of up to $0.5$ $R_{\odot}$ in the 3D CME reconstruction.

The analysis reported here also demonstrates that halo CME can be well characterized unless their projected altitude $\rho$ is too small. In particular, even if it is possible to provide images of the inner part of the solar corona ($\rho < 1.4$ R$_\odot$ usually unobserved by externally occulted coronagraphs), the visible light emission of halo CMEs expanding at these low projected altitudes $\rho$ cannot be ever used for reliable 3D reconstructions with the polarization technique because of the very large errors described above (blue regions in top panels of Figure \ref{fig:fig01}). Moreover, as the blob propagates into the corona, even when not considering any blob expansion (hence any blob density decrease), its brightness decreases as well because of the decrease in the solid angle subtended by the solar disk. Because of Thomson scattering geometry, the emitted $pB$ decreases even faster than the emitted $tB$ and this quikly makes the detection of $pB/uB$ ratio difficult as the CME starts to be farther than $\sim 5$ R$_\odot$ from the POS for low projected distances. This point is clearly shown in Figure \ref{fig:fig05}: as the blob moves away from the POS the $pB/tB$ ratio progressively decreases and it is very small almost everywhere for small LOS projected distances ($\rho < 1.4$ R$_\odot$). In this region the $pB$ emitted by the blob alone decreases much more quikly than its $tB$ emission as it is moving away from the POS (Figure \ref{fig:fig05}, top left and middle panels). As a consequence, once the background corona is also taken into account, the $pB$ and $tB$ emissions are both dominated by the background corona itself (Figure \ref{fig:fig05}, bottom left and middle panels), thus leading to large errors in the 3D reconstruction of the LOS blob location for small LOS projected distances (blue regions in top panels of Figure \ref{fig:fig01}.

In conclusion, these results have important consequences not only for future 3D reconstruction of CMEs with the polarization ratio technique, but also for the design of future coronagraphs aimed at providing a continuous monitoring of halo-CMEs for space weather prediction purposes. First, as mentioned, we suggest excluding the H$\alpha$ emission from the instrument band-pass in order to avoid problems with the determination of 3D location of CME cores. Second, monitoring of halo-CME propagation with the polarization ratio technique will not necessarily require a coronagraph field of view extending below a heliocentric distance of $\sim 1.4$ R$_\odot$, because in all cases 3D reconstructions performed with the polarization ratio technique will be subject to very large uncertainties. Nevertheless, in the analysis presented here only a single blob with constant and Gaussian density distributions is considered along the LOS, thus possible uncertainties related to the location of multiple CME features aligned along the same LOS are not considered here. In a future work (now in preparation) we will also investigate asymmetrical LOS density distributions by applying the same analysis described here to full 3D MHD simulations of CMEs inspired to those recently developed by \citet{Pagano2013}.

\begin{acknowledgements}
P.P. would like to thank STFC for financial support.
\end{acknowledgements}

\bibliographystyle{aa}
\bibliography{ref}

\end{document}